\documentclass[%
 reprint,
 amsmath,amssymb,
 aps,
]{revtex4-2}

\usepackage{graphicx}
\usepackage{dcolumn}
\usepackage{bm}
\usepackage[usenames]{color}
\usepackage[hidelinks,colorlinks=true]{hyperref}

\begin{document}

\preprint{APS/123-QED}

\title{Higgs Boson Mass Corrections at Three-Loops in the Top-Yukawa Sector \\ of the Standard Model}

\author{E. A. Reyes R.}
 \email{eareyesro@unal.edu.co}
 \affiliation{
Departamento de Fisica, Universidad de Pamplona,\\ Pamplona - Norte de Santander, Colombia.}
\author{A. R. Fazio}%
 \email{arfazio@unal.edu.co}
\affiliation{
Departamento de Fisica, Universidad Nacional de Colombia,\\ Bogotá, Colombia. \\ }


\begin{abstract}
The search for new physics signals in Higgs precision measurements plays a pivotal role in the High-Luminosity Large Hadron Collider (HL-LHC) and future colliders programs. The Higgs properties are expected to be measured with great experimental precision, implying higher-order perturbative computations of the electroweak parameters from the theoretical side. In particular, the renormalized Higgs boson mass parameter in the Standard Model shows significant variation around the electroweak scale, resulting in a lower-bound theoretical uncertainty that exceeds future collider expectations. A more stable result under the renormalization group can be computed from a non-zero external momentum Higgs self-energy, for which available calculations include 3-loop corrections in the QCD sector. In this work, we present an additional contribution by estimating the leading non-QCD 3-loop corrections to the mass of the Higgs boson in the top-Yukawa sector of order $y_t^6$. The momentum-dependent Higgs self-energy is computed in the tadpole-free scheme for the Higgs vacuum expectation value in the Landau gauge, and the explicit dependence upon the Higgs boson and top quark masses is shown. The obtained result is expressed in dimensional regularization as a superposition of a set of master integrals with coefficients that are free of poles in four space-time dimensions, and the corrections are evaluated numerically by the sector decomposition method.     

\end{abstract}

\maketitle


\section{\label{sec:Introduction} Introduction \protect}

The experiments have recently shown that high-precision measurements of the observables in the electroweak (EW) sector of the Standard Model (SM) are diverging from the theoretical predictions. In the past year, the Fermilab MUON g-2 collaboration~\cite{B.Abi} published its results concerning the muon anomalous magnetic moment, revealing a $4.2\sigma$ deviation between the experimental value and the SM predictions. Recently, another EW observable joins this list of anomalous measurements, namely the mass of the $W$-boson. The CDF collaboration~\cite{CDF} reported a new and more precise value, $M_W=80433.5 \pm 9.4 \, MeV$, together with the complete dataset collected by the CDF II detector at the Fermilab Tevatron. The current SM prediction exhibits a $7\sigma$ tension compared with the CDF measurement, suggesting the possibility to improve the SM calculations or to extend the SM. New and more precise experiments can help to explain the origin of those discrepancies, but this requires also an improvement on the precision of the theoretical calculations. Particularly, the Higgs boson mass is one of the observables to be measured with greater accuracy at future accelerators, hence an enhancement in the theoretical prediction within the SM will also be required. Although the mass of the Higgs boson in the SM is a free parameter that is adjusted to match experimental measurements, its theoretical calculation holds significant importance in estimating radiative corrections of electroweak precision observables, such as Higgs decay rates, Higgs production cross-sections, and Higgs couplings~\cite{Zerwas, HCSGroup, FreitasPoS}. These theoretical calculations provide essential insights for validating and advancing the SM in accordance with experimental Higgs data. Furthermore, the precise relationships between the Higgs mass and the SM parameters are crucial for verifying the internal consistency of the theory. For instance, they play an important role in studying the stability of the Higgs potential and determining the status of the SM vacuum~\cite{Degrassi, Buttazzo}. \\ Improvements in the theoretical calculation of the Higgs mass can be achieved by computing higher-order corrections that were previously neglected due to kinematic constraints or truncation of perturbative expansions at a certain level. In the SM, the truncation is done at three-loop order. The one- and two-loop level corrections to the Higgs self-energy have been completely computed~\cite{Kniehl1, Martin1, Kniehl2} and implemented in the public computer codes \texttt{mr}~\cite{mr} and \texttt{SMDR}~\cite{SMDR}. Additionally, the 3-loop corrections that account for the external momentum dependence of terms proportional to $g_s^4 y_t^2 M_t^2$ were evaluated in~\cite{Martin2,Martin3}, assuming the EW gaugeless limit where the contributions proportional to the SU(2) gauge couplings are disregarded. Moreover, the 1PI effective potential was used in the latter references to calculate the 3-loop contributions proportional to $g_s^2 y_t^4 M_t^2$ and $y_t^6 M_t^2$, from which the 1PI self-energies at vanishing external momenta can be derived. All these 3-loop corrections have been implemented in the latest version of the code \texttt{SMDR}. \\ The mass of the Higgs boson is not uniquely defined because it depends on the precise definition of the vacuum expectation value (vev) of the Higgs field. In the \texttt{mr} code scheme, the renormalized vev of the Higgs field is defined as the minimum of the tree-level Higgs potential. Consequently, the corrections to the mass parameters are gauge invariant due to the explicit insertion of the tadpole diagrams. However, this approach has a drawback as the Higgs tadpoles may contain negative powers of the Higgs quartic self-coupling, leading to substantial corrections in $\overline{MS}$ schemes that undermine the perturbative stability. On the other hand, the corrections included in \texttt{SMDR} typically lead to stable perturbative predictions but suffer from gauge dependences since the vacuum is defined as the minimum of the Higgs effective potential. Therefore, the tadpoles are removed by imposing an appropriate renormalization condition. Previous works have highlighted the importance of obtaining a gauge-independent prediction for the Higgs mass with stable perturbative behavior. However, a prescription that properly accounts for tadpole contributions in EW renormalization has only been developed up to the 1-loop order in~\cite{Heidi1, Heidi2}. The \texttt{SMDR} code provides the most precise predictions of the Higgs boson mass to date. Although it exhibits a renormalization scale dependence of several tens of MeV, implying theoretical uncertainties larger than the expected experimental uncertainties of approximately $10$-$20$ MeV for the upcoming HL-LHC, ILC, and FCC-ee experiments~\cite{Blas}. A more accurate calculation is therefore necessary, which includes the missing three- and even 4-loop corrections. \\ In this paper, we present an additional contribution to the ongoing efforts to compute the higher-order perturbative corrections to the Higgs boson mass. Specifically, we determine the 3-loop Higgs self-energy corrections at order $y_t^6$, arising from the non-QCD top-Yukawa sector of the SM, using the same tadpole-free prescription for the Higgs vev as that of the \texttt{SMDR} code. These 3-loop corrections are intended to be included in the calculation of the physical Higgs boson mass ($M_h$), which is obtained from the complex pole of the Higgs propagator in an on-shell scheme. Therefore, the Higgs self-energies are evaluated at non-vanishing external momentum, $p^\mu\neq 0$. Since the ratio $M_h^2/M_t^2 \approx 0.5 - 0.6$ may not be a really small expansion parameter, considering the large uncertainty in the mass of the top quark, the leading 3-loop corrections may receive significant contributions from the external momentum-dependent terms evaluated at $p^2=M_h^2$. Additionally, the inclusion of the non-vanishing external momentum self-energies is expected to cancel the renormalization scale dependence introduced in the propagator pole by the running Higgs mass computed in the effective potential approach~\cite{Quiroz1,Quiroz2}. \\ Finally, we point out that electroweak contributions at the 3-loop level are still missing. However, the analytic results for all master integrals contributing to the 3-loop Higgs self-energy diagrams in the mixed EW-QCD sector at order $\alpha^2\alpha_s$, including terms proportional to the product of the bottom and top Yukawa couplings, $y_by_t$, have been presented in~\cite{Weinzierl}. Moreover, additional identities satisfied by 3-loop self-energy Master Integrals (MIs) with four and five propagators, which enable a straightforward numerical evaluation for a generic configuration of the masses in the propagators, have been recently reported in~\cite{Martin4}. \\ The paper is organized as follows. In Section~\ref{sec:Selfenergies}, we provide technical details about the generation and regularization of the amplitudes for the 3-loop Higgs self-energies involved in our calculation. In Section~\ref{sec:Masters}, we present a Feynman integral reduction procedure and discuss the selection of a good basis of master integrals. A numerical analysis, evaluating the obtained 3-loop corrections to the Higgs mass at O($y_t^6$) as a function of the renormalization scale, is presented in Section~\ref{sec:Numerical}. Finally, we provide our conclusions and an outlook for further research in Section~\ref{sec:Conclusions}.      

\section{\label{sec:Selfenergies} Regularized Higgs Self-energies \protect}
\begin{figure}
    \centering
    \includegraphics[width=0.45\textwidth]{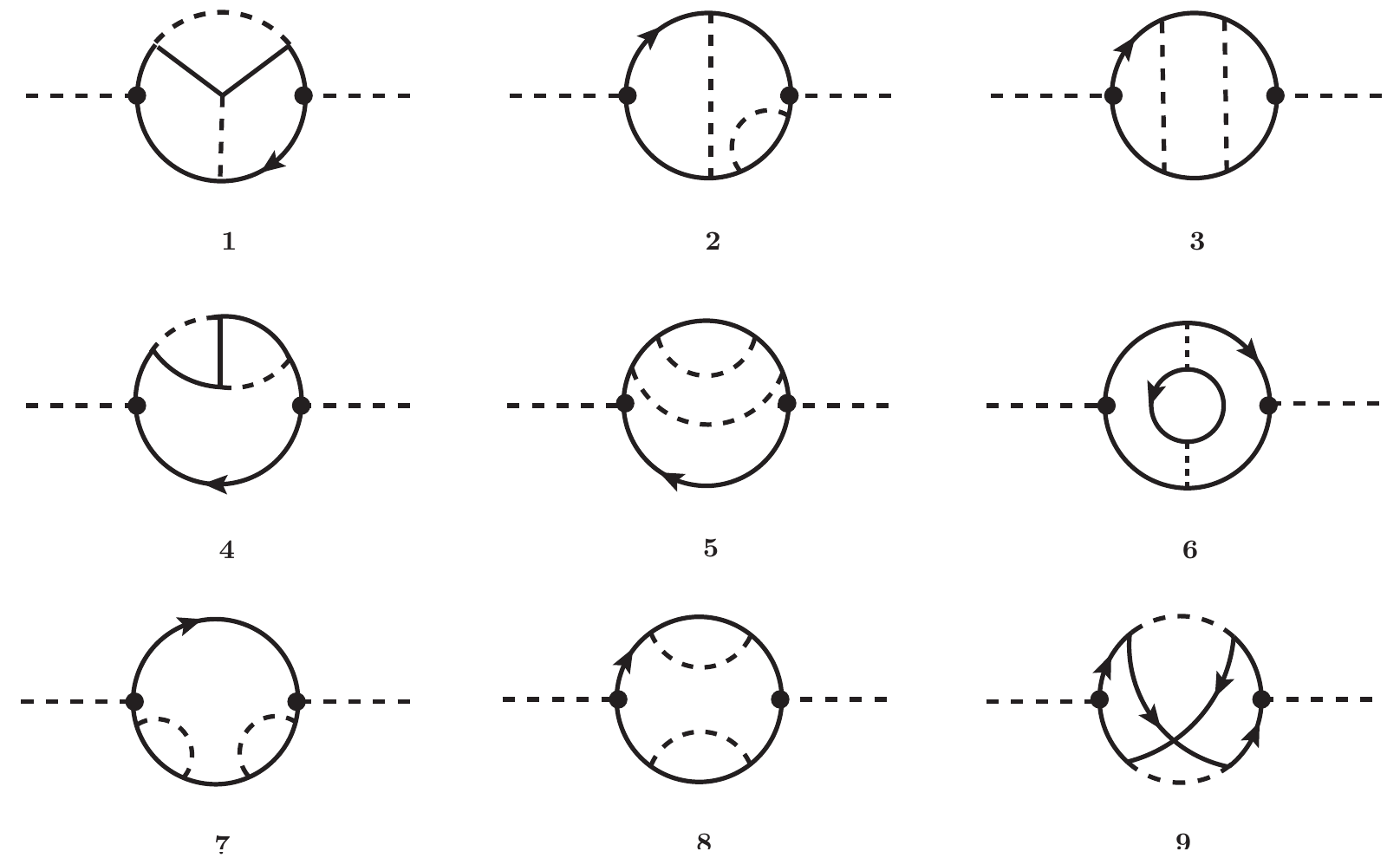}
    \caption{Examples of diagrams contributing to the O($y_t^6$) Higgs self-energy corrections in the non-QCD sector. The external dashed lines represent the Higgs boson field. The internal dashed lines represent all possible contributing scalar fields, while the solid lines represent a top or a bottom quark. Only propagators with a top quark line are massive.}
    \label{fig:topologies}
\end{figure}

In this work, our focus has been on the contributions arising from the 3-loop self-energy corrections to the Higgs boson mass, taking into account the external momentum dependence. The Higgs self-energies have been computed at order $y_{t}^{6}$ in the non-QCD sector of the SM. Thus, we have assumed the non-light fermion limit and therefore disregarded the Yukawa couplings and masses of other fermions compared to those of the top quark. The complete expression is written as 
\begin{align}
\Pi_{hh}^{(3,\, y_t^6)} & =y_{t}^{6}(\Delta_{0}+t\Delta_{1}+t^{2}\Delta_{2}+t^{3}\Delta_{3}) \nonumber \\
 & \quad +\:s\,y_{t}^{6}(\Delta_{0}^{s}+t\Delta_{1}^{s}+t^{2}\Delta_{2}^{s}),
\end{align}
where $t$ represents the squared top mass, $t=M_t^2$, while $s$ stands for the squared momentum in the external lines of the Higgs self-energies, $s=p^2$. \\ In order to obtain the expressions of $\Delta_i$ and $\Delta_j^s$ it is necessary to generate the Higgs self-energy diagrams and their corresponding amplitudes. This has been done with the help of the \texttt{Mathematica} package \texttt{FeynArts}~\cite{FeynArts1,FeynArts2}. At the considered perturbative order, only the nine different self-energy topologies depicted in FIG.~\ref{fig:topologies} contribute. Note that topologies consisting solely of cubic vertices are required. To generate such diagrams, it is necessary to set the 'adjacency' option to 3 in the \texttt{CreateTopologies} function of \texttt{FeynArts}. 
Moreover, the computation was done in a scheme where the renormalized vacuum expectation value of the Higgs field is taken to be the minimum of the Higgs effective potential. As a result, the self-energies are computed using 1PI diagrams that do not contain tadpole insertions, as discussed in references~\cite{Heidi1, Bohm, Denner, Edilson1}. Although this scheme is known to be numerically stable as it excludes terms with negative powers of the Higgs self-coupling, it has the drawback that self-energies become gauge-dependent quantities. In this work, we have adopted the Landau gauge, where the Goldstone bosons are massless, in order to minimize the number of energy scales appearing in the Feynman integrals. \\ Once the particle content is included in the nine topologies using the \texttt{InsertFields} function of \texttt{FeynArts}, the number of generated self-energy diagrams that have non-zero amplitudes at $y_t^6$ increases to 125. Examples of such diagrams are also shown in FIG.~\ref{fig:topologies}. Note that the external dashed lines represent the propagation of the Higgs field ($h$), while the internal lines, in the non-light fermions limit of the non-QCD sector, can propagate fermions (solid lines) such as the top quark ($t$) and bottom quark ($b$) fields, as well as scalars like the Higgs and the Goldstone bosons ($G^{0}$ and $G^{\pm}$) fields. The cubic vertices involved in these computation include $hht$, $G^{0}G^{0}t$ and $G^{\pm}tb$. The contribution of the bottom mass to the latter vertex is disregarded when it appears in the numerators of the integrands.\\
The considered 3-loop self-energy integrals are ultraviolet divergent in four-dimensions since all of them contain two scalar and six fermionic propagators; therefore, they are analytically continued to $D=4-2\varepsilon$ dimensions using the dimensional regularization (DREG) scheme~\cite{DREG1,DREG2,DREG3,DREG4}. In order to implement the regularization prescription, the \texttt{FeynArts} amplitudes are exported to the language of \texttt{FeynCalc}~\cite{FeynCalc1,FeynCalc2}, which is a \texttt{Mathematica} code designed for algebraic manipulations in multi-loop Feynman integral calculations. The gamma matrices are defined as a set of $D$ matrices satisfying the relation
\begin{equation}
    \label{eq:Clifford}
    \left\{ \gamma^{\mu}, \gamma^{\nu} \right\} = 2g^{\mu \nu}I; \qquad \mathrm{Tr}I = 4.  
\end{equation}
Feynman diagrams involving the charged Goldstone bosons, $G^{\pm}$, can introduce complications when dealing with traces that involve $\gamma_5$ and an arbitrary number of gamma matrices. In such cases, we employ the practical non-cyclicity prescription~\cite{korner,Jegerlehner}, which imposes certain conditions. Specifically, $\gamma_5$ is treated as an anticommuting object that satisfies
\begin{eqnarray}
\left\{ \gamma_{5},\:\gamma^{\mu}\right\} =0;\qquad \gamma_{5}^{2}=1,
\end{eqnarray}
and it is forbidden to use cyclicity in traces involving an odd number of $\gamma_5$ matrices. Using the above anticommutation relation and the Clifford algebra in eq.~(\ref{eq:Clifford}), it is possible to order any product of Dirac matrices in a canonical way. In particular, when there is an even number of $\gamma_{5}$ matrices, they can be entirely eliminated. However, in the case of an odd number, a single $\gamma_{5}$ always remains and is consistently moved to the right of the product. It is worth noting that the diagrams under consideration involve four independent momentum scales: the external momentum $p$ and the loop momenta $q_{1}$, $q_{2}$, and $q_{3}$. As a result, these diagrams may include traces with a single $\gamma_{5}$ and, at most, four $\gamma$ matrices. Consequently, the following relationships are also necessary:
\begin{equation}
\label{eq:oddg}
\mathrm{Tr}\left[\gamma_{5}\right] = \mathrm{Tr}\left[\gamma^{\mu_1}\dots\gamma^{\mu_{2n-1}}\gamma_{5}\right]=0,    
\end{equation}
\begin{equation}
\label{eq:4g}
\mathrm{Tr}\left[\prod_{j=1}^4\gamma^{\mu_j}\gamma_{5}\right]=
\begin{cases}
-4i\epsilon^{\mu_1\mu_2\mu_3\mu_4} & \mu_j \in\{0,1,2,3\}\\
0 & \mu_j \in\{4,\dots,D-1\}
\end{cases},
\end{equation}
where $\epsilon^{\mu_1\mu_2\mu_3\mu_4}$ represents the four-dimensional Levi-Civita tensor. Note that in this non-cyclicity approach, $\gamma_5$ odd traces are considered as inherently four-dimensional quantities. A further examination of all the Feynman diagrams for each topology in FIG.\ref{fig:topologies} reveals that topologies 1, 4, 6, and 9 do not involve traces with the matrix $\gamma_5$. For topologies 5 and 8, the traces with one $\gamma_5$ and at most three $\gamma$ matrices vanish as given by eq.(\ref{eq:oddg}). In the case of topologies 2 and 7, when the amplitudes of each topology are summed, there is a cancellation of terms associated with any trace involving the matrix $\gamma_5$. Finally, topology 3 contains contributions with a trace of a single $\gamma_5$ and four $\gamma$ matrices, which need to be evaluated according to eq.~(\ref{eq:4g}). \\
In addition, it is worth mentioning that for amplitudes with closed fermion-loops, which is the case of all the topologies in FIG.~\ref{fig:topologies}, both the usual Breitenlohner-Maison scheme~\cite{BM0,BM} and the non-cyclicity scheme considered in our calculation yield identical results. 

\section{\label{sec:Masters} Good Master Integrals}

Once the amplitudes are regularized, each of them can be expressed as a superposition of a large set of about one thousand integrals with the following structure: 
\begin{align}
\left\langle \frac{\mathcal{N}(q_{i} \cdot q_{j},q_{i}\cdot p,p^2)}{D_{1}^{\nu_{1}}D_{2}^{\nu_{2}}D_{3}^{\nu_{3}}D_{4}^{\nu_{4}}D_{5}^{\nu_{5}}D_{6}^{\nu_{6}}D_{7}^{\nu_{7}}D_{8}^{\nu_{8}}D_{9}^{\nu_{9}}D_{0}^{\nu_{0}}}\right\rangle _{3l},\\
\left\langle \left(\dots\right)\right\rangle _{3l}=(Q^2)^{3\varepsilon}\int \frac{d^{D}q_{1}}{(2\pi)^D}\int \frac{d^{D}q_{2}}{{(2\pi)^D}}\int \frac{d^{D}q_{3}}{{(2\pi)^D}}, \nonumber
\end{align}
where $Q$ is the renormalization scale defined in the $\overline{MS}$ scheme,
$
    Q^2 = 4\pi e^{-\gamma_E} \mu^2,
$
in terms of the unit mass $\mu$ and of the Euler-Mascheroni constant $\gamma_E$. The denominators $D_j$ are inverse scalar propagators, denoted as:
\begin{equation}
{\small \begin{array}{cc}
D_{1}=\left(q_{1}^{2}-m_{1}^{2}\right), & D_{2}=\left(q_{2}^{2}-m_{2}^{2}\right), \\
D_{3}=\left(q_{3}^{2}-m_{3}^{2}\right), & D_{4}=\left((q_{1}-q_{2})^{2}-m_{4}^{2}\right),\\
D_{5}=\left((q_{1}-q_{3})^{2}-m_{5}^{2}\right), & D_{6}=\left((q_{2}-q_{3})^{2}-m_{6}^{2}\right),\\
D_{7}=\left((q_{1}+p)^{2}-m_{7}^{2}\right), & D_{8}=\left((q_{2}+p)^{2}-m_{8}^{2}\right),\\
D_{9}=\left((q_{3}+p)^{2}-m_{9}^{2}\right), & D_{0}=\left((q_{1}-q_{2}+q_{3}+p)^{2}-m_{0}^{2}\right), 
\end{array} }
\end{equation}
whereas the numerator $\mathcal{N}$ is a function of scalar products involving the three loop momenta and the external momentum. The coefficients of the integrals now depend on $y_t$, $t$, and $s$, while the masses in the propagators $D_j^{-1}$ can take values $m_j=0$ or $M_t$. The specific arrangement of the masses determines the family to which the integrals belong, while the set of exponents ${\nu_j}$ defines sectors within these families.
\begin{table}
\centering%
\begin{tabular}{|c|c|}
\hline 
Topology & Propagator\tabularnewline
\hline 
\hline 
1 & \{134679\}\tabularnewline
\hline 
2 & \{1278\}, \{12378\}\tabularnewline
\hline 
3 & \{1379\}, \{123789\}, \{134679\}\tabularnewline
\hline 
4 & \{24589\}\tabularnewline
\hline 
5 & \{258\}, \{278\}, \{2578\}, \{24589\}\tabularnewline
\hline 
6 & \{125678\}\tabularnewline
\hline 
7 & \{17\}, \{147\}, \{157\}, \{1457\}\tabularnewline
\hline 
8 & \{17\}, \{127\}, \{157\}, \{1257\}\tabularnewline
\hline 
9 & \{123790\}\tabularnewline
\hline 
\end{tabular}

\caption{Integral families are represented with a list \{$ijk...$\}, where each number in the list indicates the position ``$j$" of a massive propagator $D_j^{-1}$. The propagators not included in the list are massless.} \label{tab:IntFam}

\end{table}
For the planar diagrams, represented by topologies 1 to 8, the denominator $D_0$ is removed by setting $\nu_0=0$. For the non-planar diagrams in topology 9, we have $\nu_8=0$. It is important to note that in order to express any scalar product in $\mathcal{N}$ as a combination of inverse propagators, a basis of nine propagators is needed for each family. As a result, the numerator $\mathcal{N}$ is rewritten in terms of the $D_j$'s, leading to scalar integrals that may also include irreducible numerators, which are denominators with negative integer exponents. The resulting integral families for each topology are listed in Table~\ref{tab:IntFam}. It should be noted that an individual topology can contain multiple families, and each family can include up to six massive propagators. Additionally, the exponents {$\nu_j$} range from $-3$ to $3$. \\ The obtained set of scalar integrals is not mutually independent, but rather connected through integration by parts (IBP) and Lorentz Invariant (LI) identities. These relations allow us to express any scalar integral as a linear combination of a basis of Master Integrals, denoted as
\begin{equation}
    \tilde{G}_{\{\nu_0,\dots,\nu_9\}} = \left\langle \prod_{j=0}^{9}D_{j}^{-\nu_{j}} \right\rangle _{3l}.
\end{equation}
The coefficients in this linear combination are rational functions of polynomials that depend on the space-time dimension and all the kinematical invariants involved in the calculation. To handle this reduction process, we utilized the code \texttt{Reduze}~\cite{Reduze, Reduze2}, which performs the reduction of scalar integrals to a set of basis integrals. As expected, in complex scenarios such as the IBP reduction of 3-loop self-energy integrals involving multiple energy scales, the basis provided by \texttt{Reduze} can be inefficient. This inefficiency arises from the cumbersome denominators of some coefficients in the MIs, which contain large expressions that require significant processing time and memory resources. Furthermore, these denominators may also include kinematical singularities (independent of $D$), described by the Landau conditions~\cite{Landau}, and/or divergences in $D-4=2\varepsilon$ (independent of the kinematical invariants), which require evaluating the finite parts of the Laurent expansion in $\varepsilon$ of the MIs~\cite{Binoth, Bogner, FIESTA1}. To address this situation, we employ the prescription discussed in~\cite{Smirnovs}, which is based on Sabbah's theorem~\cite{Sabbah}. We have implemented a transition from the ``bad" basis of Master Integrals (MIs) to an appropriate basis, denoted as $G_{{j}}$, using \texttt{Mathematica} and the assistance of \texttt{FIRE}~\cite{FIRE5, FIRE6}. This transformation ensures that the denominators of the coefficients in the new basis are simpler expressions that are devoid of kinematical and non-kinematical singularities. Thus, the selection of the new master integrals has been made by ensuring that the polynomials in the denominators of the coefficients do not vanish in the limit where $D-4$ approaches zero. The Sabbah's theorem guarantees the existence of such a ``good" basis, but in practice, this implies finding additional relations between the master integrals, such that
\begin{equation}
    \tilde{G}_{\{i\}} = \sum_{j=1}^{|\sigma|} \frac{n_{i,j}}{d_{i,j}} G_{\{j\}}, 
\end{equation}
for a given sector $\sigma$ of which $|\sigma|$ represents the number of integrals in the corresponding sector. In these relations, the coefficients $n_{i,j}$ must include products of polynomials that cancel out the problematic denominators of the coefficients of the original master integrals $\tilde{G}_{\{i\}}$ in the initial IBP reduction. On the other hand, the coefficients $d_{i,j}$ must be chosen as good denominators free of singularities. A simple example can be found in the family \{134679\} of the first topology (see FIG.~\ref{fig:topologies} and Table~\ref{tab:IntFam}). A bad election of the basis in the reduction procedure can lead to coefficients with null denominators for $D=4$, of the form 
{\small\begin{align}
(-5+D)(-4+D)(-3+D)(-10+3D)st^{2}(-s+2t) \nonumber \\
\times (-s+4t)(-s+10t)(s^{2}-16st+24t^{2})
\end{align}}or an even worse coefficient can arise with denominator 
{\small\begin{align}  
2(-4+D)(s-4t)^{2}t(-38997504s^{18}+159422976Ds^{18}) \nonumber \\
\times t(-288550464D^{2}s^{18}+\dots+244~\text{terms}),
\label{coeff2}
\end{align}}manifesting moreover threshold singularities. The denominator of eq.~(\ref{coeff2}) arises from the sector involving the master integrals $\tilde{G}_{\{-1, 0, 1, 1, 0, 1, 1, 0, 0\}}$, $\tilde{G}_{\{0, 0, 2, 1, 0, 2, 1, 0, 0\}}$ and $\tilde{G}_{\{0, 0, 1, 1, 0, 1, 1, 0, 0\}}$. However, by choosing a better basis comprising the master integrals $G_{\{1, -1, 1, 1, 1, 1, 1, 1, 0\}}$, $G_{\{1, 0, 1, 1, 1, 1, 1, 1, 1\}}$, $G_{\{0, 0, 1, 1, 1, 1, 1, 1, 1\}}$, we can avoid this issue and obtain a simpler expression for the total amplitude of the first topology: 
\begin{align}
\mathcal{A}_{1}^{^{\{134679\}}} & =y_{t}^{6}\left[t \, \left(4G_{_{\{0,0,1,1,1,0,1,1,1\}}}+2G_{_{\{0,0,1,1,1,1,0,1,1\}}}\right.\right. \nonumber \\
 & \qquad\quad-4G_{_{\{0,0,1,1,1,1,1,0,1\}}}-4G_{_{\{1,-1,1,1,0,1,1,1,1\}}} \nonumber \\
 & \qquad\quad+2G_{_{\{1,-1,1,1,1,1,0,1,1\}}}+2G_{_{\{1,-1,1,1,1,1,1,1,0\}}} \nonumber \\
 & \qquad\quad+4G_{_{\{1,0,0,0,1,1,1,1,1\}}}-4G_{_{\{1,0,0,1,1,1,1,0,1\}}} \nonumber \\
 & \qquad\quad+2G_{_{\{1,0,0,1,1,1,1,1,0\}}}-4G_{_{\{1,0,1,1,0,1,0,1,1\}}} \nonumber \\
 & \qquad\quad+4G_{_{\{1,0,1,1,0,1,1,0,1\}}}-4G_{_{\{1,0,1,1,0,1,1,1,0\}}} \nonumber \\
 & \qquad\quad+2G_{_{\{1,0,1,1,1,1,-1,1,1\}}}-2G_{_{\{1,0,1,1,1,1,0,0,1\}}} \nonumber \\
 & \qquad\quad\left.-2G_{_{\{1,0,1,1,1,1,1,0,0\}}}+2G_{_{\{1,0,1,1,1,1,1,1,-1\}}}\right) \nonumber \\
 & \qquad + t^{2}\, \left(8G_{_{\{0,0,1,1,1,1,1,1,1\}}}+8G_{_{\{1,0,0,1,1,1,1,1,1\}}}\right. \nonumber \\
 & \qquad\quad+8G_{_{\{1,0,1,0,1,1,1,1,1\}}}-16G_{_{\{1,0,1,1,0,1,1,1,1\}}} \nonumber \\
 & \qquad\quad+8G_{_{\{1,0,1,1,1,0,1,1,1\}}}+8G_{_{\{1,0,1,1,1,1,0,1,1\}}} \nonumber \\
 & \qquad\quad\left.-16G_{_{\{1,0,1,1,1,1,1,0,1\}}}+8G_{_{\{1,0,1,1,1,1,1,1,0\}}}\right) \nonumber \\
 & \qquad + \left.t^{3}\, 32G_{_{\{1,0,1,1,1,1,1,1,1\}}}\right] \nonumber \\
 & +sy_{t}^{6}\left[t\,\left(-2G_{_{\{1,0,1,0,1,1,1,1,1\}}}+4G_{_{\{1,0,1,1,0,1,1,1,1\}}}\right.\right. \nonumber \\
 & \qquad\quad-2G_{_{\{1,0,1,1,1,0,1,1,1\}}}-2G_{_{\{1,0,1,1,1,1,0,1,1\}}} \nonumber \\
 & \qquad\quad\left.+4G_{_{\{1,0,1,1,1,1,1,0,1\}}}-2G_{_{\{1,0,1,1,1,1,1,1,0\}}}\right)\nonumber \\
 & \qquad \left.-t^{2} \, \, 8G_{_{\{1,0,1,1,1,1,1,1,1\}}}\right].
\end{align}
It should be noted that the master integrals contain 9 indices due to the omission of $D_0$ in the planar topologies, while $D_8$ is removed in non-planar diagrams. Analogous simple expressions have also been derived for topologies 2, 4 and 6, while the results for the amplitudes $\mathcal{A}_3$, $\mathcal{A}_5$, $\mathcal{A}_7$, $\mathcal{A}_8$ and $\mathcal{A}_9$ are more extensive. All amplitudes, along with a list of good master integrals, useful IBP reductions, and the main \texttt{Mathematica} routines utilized in this computation, can be accessed through the following link: \url{https://github.com/fisicateoricaUDP/HiggsSM}. In particular, the planar diagrams can be reduced to a superposition of 212 MIs, while the non-planar diagrams can be expressed in terms of 82 masters. Even if a good basis of MIs could be found with the help of Sabbah's theorem in this computation, when an additional energy scale is included, such as the mass of the bottom quark, the coefficients of the obtained master integrals become even worse and render any IBP reduction procedure inefficient. This kind of problem also arises in beyond the Standard Model theories, such as in the calculation of $M_h$ in SUSY, where the analogous contribution at order $y_t^6$ is missing~\cite{HiggsSUSY}. In this case, additional scales, like the squark masses, need to be considered. Analytical approaches, such as the Loop-Tree Duality technique~\cite{LTD1,LTD2}, can be employed as an interesting alternative for directly evaluating the amplitudes in scenarios involving multiple energy scales.

\section{\label{sec:Numerical} Numerical Analysis}

\begin{figure}
    \centering
    \includegraphics[width=0.47\textwidth]{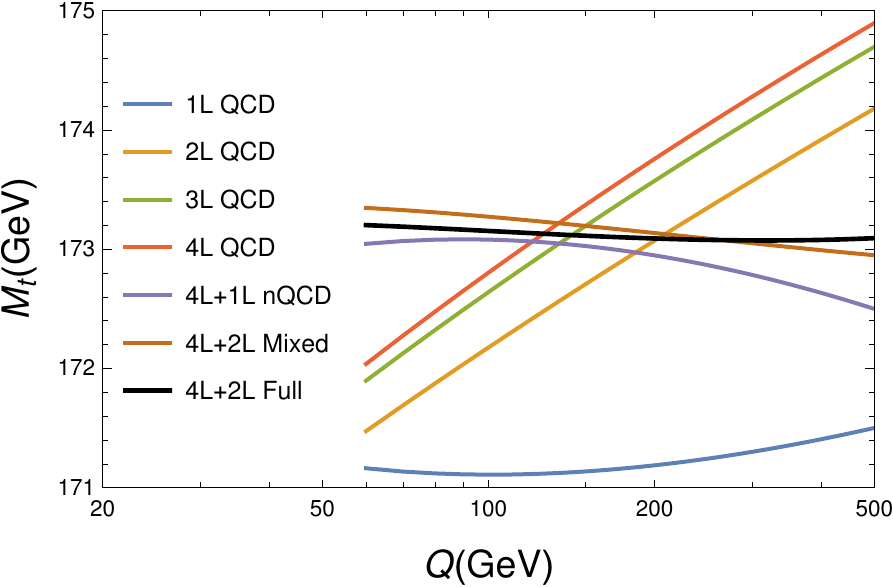}
    \caption{Evolution of the top quark mass $M_t$ as a function of the renormalization scale $Q$ in the $\overline{MS}$ scheme. The different perturbative contributions are shown. In particular, the black line contains the 4-loop QCD and the full 2-loop EW corrections.}
    \label{fig:PlotMt}
\end{figure}

\begin{figure}
    \centering
    \includegraphics[width=0.47\textwidth]{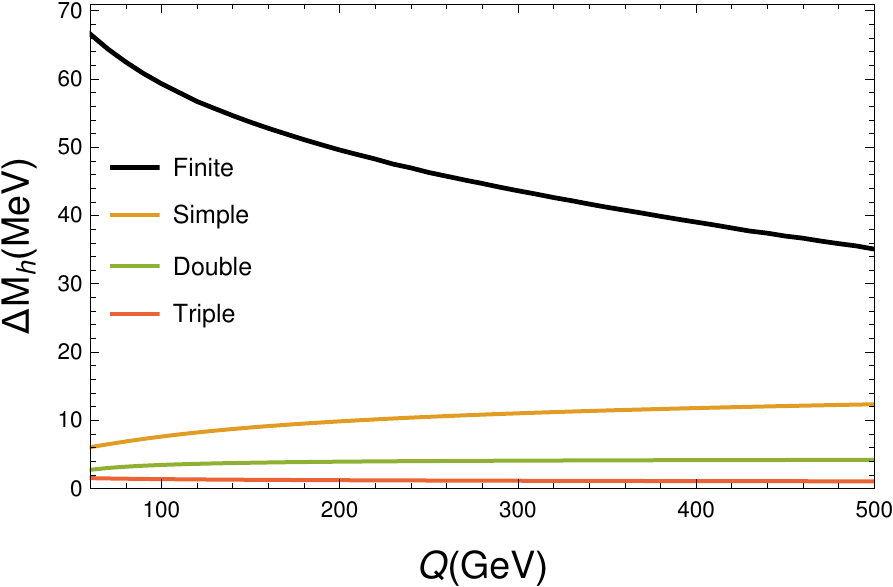}
    \caption{Renormalization group scale dependence coming from the external momentum contribution to the 3-loop Higgs self-energy correction at order $y_t^6$ in the SM. The evolution of the finite part and the coefficients of the simple, double and triple poles have been included.}
    \label{fig:PlotDMh}
\end{figure}

In this section, we present an estimate of the numerical value of the complete 3-loop 1PI Higgs self-energy contribution at order $y_t^6$. To obtain the resulting expression, we sum all the amplitudes $\mathcal{A}_j$ from the 21 families reported in Table~\ref{tab:IntFam},
\begin{equation}
    \Pi_{hh}^{(3,\, y_t^6)}(s,Q, M_t, y_t) = \sum_j \mathcal{A}_j. \label{sigmahh}
\end{equation}
The Higgs self-energy in eq.~(\ref{sigmahh}) is intended to contribute to the calculation of the complex pole of the Higgs propagator. The real part of this pole, which corresponds to the physical mass of the Higgs boson, $M_h=125.1~\mathrm{GeV}$~\cite{MhExp}, is obtained by solving the equation:
\begin{equation}
    M_h^2 -i\Gamma_h M_h = m_B^2 + 3\lambda_B v_B^2 + \sum_{l=1}^{3} \frac{1}{(16\pi^4)^{l}} \Pi_{Bhh}^{(l)}(s), \label{eq:spole}
\end{equation}
where
\begin{equation}
\Pi_{Bhh}^{(3)}(s) = \Pi_{Bhh}^{(3,g_s^4 y_t^2)}(s) + \Pi_{Bhh}^{(3, g_s^2 y_t^4)}(s) + \Pi_{Bhh}^{(3, y_t^6)}(s). \label{PiBhh(3)}    
\end{equation}
The effects of the 3-loop order in the EW sector, as well as higher-order terms, are not included in eq.~(\ref{PiBhh(3)}) as they have not been computed by any research group to date. The mass parameter $m_B^2$, the self-interacting Higgs quartic coupling $\lambda_B$, the Higgs vev $v_B$ and the parameters included in the self-energy functions $\Pi_{Bhh}^{(l)}$ are all bare quantities. These bare parameters can be related to their corresponding $\overline{MS}$ renormalized counterparts through relations of the form 
\begin{align}
    \chi_B = \chi + \sum_{l}\frac{1}{(16\pi^4)^{l}} \delta^{(l)}\chi, \label{MSbarrel}
\end{align}
where $\chi = m^2, \lambda, v^2, y_t, \dots $ are $\overline{MS}$ parameters. The main advantage of performing the calculations in terms of bare parameters and subsequently expressing the results in terms of $\overline{MS}$ quantities using relations (\ref{MSbarrel}) is that the renormalization of the sub-divergences in the Higgs self-energies is accomplished without the need to include counter-term diagrams separately. For a 3-loop estimation of the physical Higgs boson mass, the $\overline{MS}$ relations (\ref{MSbarrel}) need to be expanded up to 3-loop order. The counter-terms, denoted as $\delta^{(l)}\chi$, can be obtained from the 3-loop beta functions and anomalous dimensions provided in~\cite{Beta3L1,Beta3L2,Beta3L3,Beta3L4,Beta3L5,Beta3L6,Beta3L7,Beta3L8}, following the procedure explained in~\cite{3LSMEP} and references therein. It is worth noting that, the difference between the 3-loop bare self-energies and the corresponding 3-loop $\overline{MS}$ self-energies arises at the 4-loop order,
\begin{equation}
         \Pi_{Bhh}^{(3)}(\chi_B) = \Pi_{hh}^{(3)}(\chi) + \frac{1}{16\pi^4}\delta^{(1)}\chi\frac{\partial}{\partial \chi} \Pi_{hh}^{(3)}(\chi) + \dots . \label{eq:4loop-div}
\end{equation}
Therefore, for our purposes, in the O($y_t^6$) Higgs self-energies we can directly substitute the bare parameters $\chi_B$ with their corresponding $\overline{MS}$ counterparts $\chi$. As a result, the parameters and amplitudes appearing in equation (\ref{sigmahh}) are quantities that can be numerically evaluated for various values of the renormalization scale $Q$ within the $\overline{MS}$ scheme.
\begin{figure}
    \centering
    \includegraphics[width=0.47\textwidth]{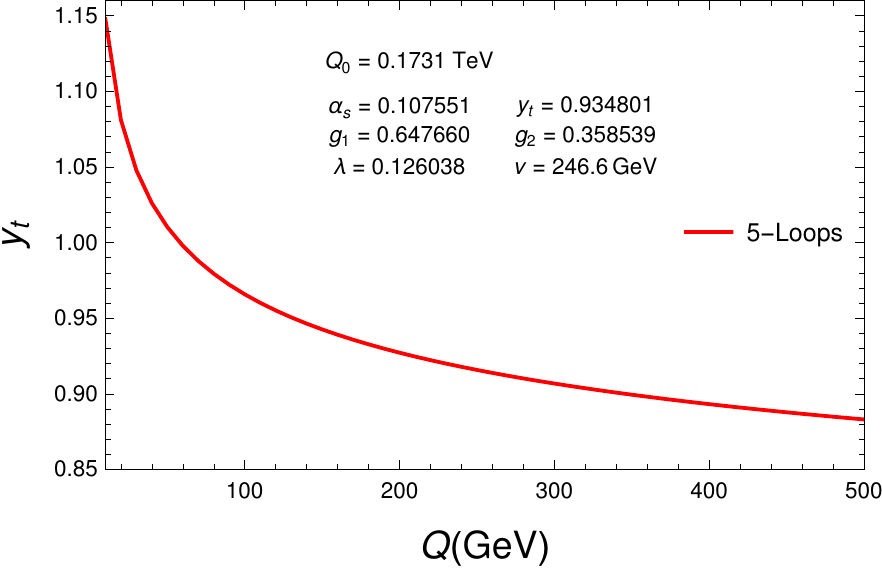}
    \caption{Renormalization group evolution of the top Yukawa coupling $y_t$ in the $\overline{MS}$ scheme including the full 3-loop RGEs for all the SM parameters and the QCD beta functions of $y_t$ and $\alpha_s$ up to 5-loops. Here $g_1$ and $g_2$ stands for the EW gauge couplings, \textsl{v} is the Higgs vev and $\lambda$ represents the quartic Higgs self-coupling.}
    \label{fig:Plotyt}
\end{figure} 
We then take the Yukawa coupling $y_t$ as an independent parameter that evolves within this framework as a function of $Q$. Meanwhile, $M_t = y_t v/\sqrt{2} + \delta M_t$ is obtained from the complex pole of the top quark propagator and is expressed in terms of the running $\overline{MS}$ parameters in the tadpole-free scheme as specified in~\cite{MtFull2L1}. We adopt this scheme because it is consistent with our choice of the renormalized vev, where the Higgs tadpoles vanish. \\ To obtain the running of $y_t$, we use the full 3-loop $\overline{MS}$ renormalization group equations (RGEs) of the SM parameters~\cite{Beta3L1,Beta3L2,Beta3L3,Beta3L4,Beta3L5,Beta3L6,Beta3L7,Beta3L8} plus the $O(\alpha_s^5)$ QCD contributions to the strong coupling beta function~\cite{Beta4L1,Beta4L2,Beta5L1,Beta5L2} and the $O(\alpha_s^5)$ QCD contributions to the beta functions of the Yukawa couplings~\cite{Beta4Lyt1,Beta4Lyt2,Beta5Lyt}. FIG.~\ref{fig:Plotyt} depicts the evolution of the top Yukawa coupling, $y_t$, as a function of the renormalization scale $Q$ in the range of $10$ to $500$~GeV. To produce this plot, we select the boundary conditions indicated at the top of the graph, that ensure the masses in the $\overline{MS}$ scheme at $Q_0=0.1731~\mathrm{TeV}$ match the experimentally reported values from the latest edition of the Review of Particle Properties~\cite{ParticleData} ($M_h=125.1$~GeV, $M_t=172.7$~GeV, etc.). We consistently employ these boundary conditions for the subsequent plots as well.\\On the other hand, the numerical values of $M_t$ as a function of the $\overline{MS}$ renormalization scale $Q$ at which it is calculated are obtained using the \texttt{SMDR} code, which employs the tadpole-free scheme for the renormalization of $M_t$. In FIG.~\ref{fig:PlotMt}, we present the results obtained through successive approximations, which include the contributions from pure QCD at 1-loop~\cite{Mt1L}, 2-loop~\cite{Mt2L}, 3-loop~\cite{Mt3L}, and 4-loop~\cite{Mt4L1, Mt4L2, Kataev:2015gvt, Kataev:2018sjv}, as well as non-QCD 1-loop, mixed EW-QCD 2-loop, and full 2-loop EW corrections to the top quark mass~\cite{MtFull2L1}. Note that the definition of the top-quark mass is not unique since the mass is proportional to the vev of the Higgs field, $v$, which can be chosen differently as discussed in Section~\ref{sec:Introduction}. The calculations of $M_t$ in various renormalization schemes can be found in~\cite{MtnonQCD1L, MtMixed2L, Kataev:2022dua}.
A discussion regarding the differences between various schemes for a running top quark mass is indeed important, as emphasized in~\cite{Kataev:2022dua}. However, such a discussion is beyond the scope of this study. In FIG.~\ref{fig:PlotMt}, it is noteworthy that the pure QCD predictions exhibit a significant scale dependence of a few GeVs as the renormalization scale $Q$ is varied from $60$ to $500$ GeV. Notably, this scale dependence is greatly reduced with the inclusion of the EW contributions. Specifically, when the full 2-loop EW contribution is added, the renormalization scale dependence decreases by approximately 97\% within the considered $Q$ range. Therefore, it is crucial to incorporate the EW corrections in any numerical analysis involving $M_t$, as emphasized in \cite{MtFull2L1, MtnonQCD1L, MtMixed2L, Kataev:2022dua}.  \\ For our preliminary numerical analysis, we utilize the data from the black curve in FIG.~\ref{fig:PlotMt}, which incorporates the complete 2-loop EW + 4-loop QCD corrections to $M_t$ in the tadpole-free scheme, as well as the data used to derive the running of $y_t$ in FIG.~\ref{fig:Plotyt}. With this data, we conduct a numerical evaluation of the O($y_t^6$) Higgs mass self-energy corrections for various kinematical points. The results for $s=0$ have already been reported in~\cite{Martin1, Martin3}, and they are consistent with our findings in the same limit. The truly new contribution of this work comes, therefore, from the terms containing the external momentum dependence, which are obtained from the difference
\begin{equation}
    \Delta M_h = \textrm{Re}\left[\Pi^{(3,\, y_t^6)}_{hh}(s\neq 0) - \Pi^{(3,\, y_t^6)}_{hh}(s = 0)\right]. \label{eq:DeltaMh}
\end{equation}
The evaluation of $\Delta M_h$ requires the calculation of 294 master integrals. These integrals were computed using the code \texttt{FIESTA 5.0}~\cite{FIESTA5}, which implements the sector decomposition approach to perform the pole resolution, a Laurent expansion in $\varepsilon = (4 - D)/2$ and numerical integration for each MI. The expansion is carried out up to order $\varepsilon^0$. Evanescent terms of order $\varepsilon^n$ with $n>0$ are not necessary since the coefficients of the good master integrals do not contain poles at $D=4$. The value of the external momentum was fixed to be equal to the physical Higgs boson mass, i.e., $s=M_h^2$. With this parameter choice, we have $s<m_j^2$, ensuring that $s$ is below any threshold and all master integrals are real. For these cases, \texttt{FIESTA} was run with specific settings. In particular, the \texttt{ComplexMode} was set to \texttt{False} and the \texttt{Precision} option was left at its default value, which corresponds to a precision of 6 digits. At a specific kinematic point, the MIs can have positive or negative numerical values, leading to a reduction in achievable numerical precision of up to 4 digits in certain cases. This reduction occurs due to cancellations between different terms in the summation of the amplitudes. Consequently, the final result has a precision of 4 digits, which is sufficient for evaluating the corrections in eq.~(\ref{eq:DeltaMh}). \\ The curves in~FIG.~\ref{fig:PlotDMh} depict the variation of $\Delta M_h$ as a function of the renormalization scale, ranging from $Q=60$~GeV to $Q=500$~GeV. Note from FIG.~\ref{fig:Plotyt} that the coupling $y_t$ exits the perturbative regime below $Q=60~\mathrm{GeV}$, and as a result, this region was excluded. The black line represents the $\overline{MS}$ renormalized finite part of $\Delta M_h$, which corresponds to the correction that should be included in the predictions of electroweak precision observables. This contribution has a magnitude of approximately $51$ MeV for $Q=173.1$ GeV and exhibits a significant dependence on the renormalization scale. Specifically, it decreases by about $47\%$ across the entire range of $Q$ values considered. In particular, when $Q$ is varied around the EW scale, from $100$~GeV to $300$~GeV, the correction is reduced by approximately $16$ MeV. Notably, this reduction is of the same order of magnitude as the anticipated experimental precision at future colliders such as HL-LHC ($10-20$ MeV \cite{HL-LHC}), ILC ($14$ MeV \cite{ILC}), and FCC-ee ($11$ MeV \cite{FCC-ee}). \\ In addition to the finite contribution, the divergent terms need to be taken into account in the Higgs propagator as described by equations (\ref{eq:spole}) and (\ref{PiBhh(3)}). However, through the reparametrization outlined in formula (\ref{MSbarrel}), the $\overline{MS}$ counter-terms effectively cancel out both the local and non-local divergences. The latter, originating from diagrams with sub-loops, are eliminated by the expansion terms
\begin{equation}
    \frac{1}{16\pi^4}\delta^{(1)}y_t^2\frac{\partial}{\partial y_t} \Pi_{hh}^{(2,y_t^4)}+ \frac{1}{(16\pi^4)^2}\delta^{(2)}y_t^2\frac{\partial}{\partial \chi} \Pi_{hh}^{(1,y_t^2)} \label{eq:sub-div}
\end{equation}
at order $y_t^6$. Note that this is equivalent to evaluating the one- and two-loop Higgs self-energies with two- and one-loop counter-term insertions, respectively. In FIG.~\ref{fig:PlotDMh}, we have included the evolution of the divergent terms, illustrating the behavior and magnitude of the contributions in eq.~(\ref{eq:sub-div}). Furthermore, these divergences play a fundamental role in the renormalization at higher orders beyond three-loops, as illustrated in eq.~(\ref{eq:4loop-div}). Specifically, the yellow, green and red lines represent the coefficients of the simple $\frac{1}{\varepsilon}$, double $\frac{1}{\varepsilon^2}$, and triple $\frac{1}{\varepsilon^3}$ poles, respectively.
The coefficients of the poles exhibit a mild dependence on the renormalization scale. The triple pole coefficient varies by approximately $0.5$ MeV for $60~\textrm{GeV} \leq Q \leq 500~\textrm{GeV}$. In this case, the dependence on $Q$ is not explicit, and the variation is primarily attributed to the renormalization group evolution of $y_t$ and $M_t$. The double pole coefficient features an explicit logarithmic dependence on $Q$, resulting in a variation of approximately $1.5$ MeV. On the other hand, the simple pole coefficient contains a squared logarithmic dependence on $Q$, leading to a variation of approximately $6.2$ MeV. \\ Finally, it is important to note that incorporating the new 3-loop corrections, denoted as $\Delta M_h$, into the complex pole mass, $s_{\textrm{pole}}^h$, of the SM Higgs boson, as well as conducting a detailed numerical analysis to assess their impact on the theoretical calculation of the Higgs boson pole mass, poses a non-trivial challenge. These computations require iterative evaluations of the MIs and amplitudes at $s=\textrm{Re}(s_{\textrm{pole}}^h)$, rather than a naive evaluation at $s=M_h^2$. The numerical determination of the Higgs boson pole mass, incorporating the pure 3-loop corrections presented in this paper, will be performed in a future phenomenological analysis.

\section{\label{sec:Conclusions} Conclusions and perspectives}
In this article, we have presented a new contribution to the perturbative corrections of the SM Higgs boson mass. This contribution arises from the pure 3-loop Higgs self-energies at order $y_t^6$, considering the external momentum dependence. The evaluation involves Feynman diagram calculations for eight planar and one non-planar topologies, all featuring only cubic vertices and a fermion loop in the internal lines. The Higgs self-energies do not include tadpole contributions, as the renormalized vev of the Higgs field is taken as the minimum of the Higgs effective potential. Consequently, the considered contributions exhibit good perturbative behavior, although with an additional gauge dependence. In our calculations, we have utilized the Landau gauge to minimize the number of energy scales present in the Feynman amplitudes. Additionally, we conducted our analysis in the gaugeless and non-light fermions limits, neglecting the masses of the electroweak vector bosons and all light fermions. As a result, the final outcome is expressed solely in terms of the top quark mass $M_t$ and the Higgs boson mass $M_h$. To regularize the Feynman amplitudes associated with the Higgs self-energies, we employed the dimensional regularization procedure. In particular, a non-ciclicity prescription was applied to handle the regularization of the $\gamma_5$ matrix. The regularized amplitudes obtained are expressed in terms of thousands of scalar integrals, which were further reduced to a superposition of a basis of master integrals using the IBP and LI identities implemented in the \texttt{Reduze} code. The automated reduction process results in a set of master integrals that exhibit large coefficients featuring kinematic singularities and non-kinematic divergences at $D=4$ space-time dimensions. As the number of scales increases, the aforementioned singular behavior and the length of the coefficient expressions become more pronounced. However, we have showed that these divergences are spurious and can be eliminated through a suitable redefinition of a well-defined basis, whose existence is ensured by Sabbah's theorem. Consequently, the expressions obtained for the amplitudes of the relevant topologies are linear combinations of 212 planar and 82 non-planar "good" master integrals, whose coefficients do not contain poles as $D$ approaches 4. One advantage of this approach is that the evanescent terms of the Laurent expansion of the master integrals are not required. A preliminary numerical analysis was conducted to determine the size of the new momentum-dependent Higgs self-energy contributions, revealing a value of approximately 51 MeV at the renormalization scale Q=0.1732 TeV. However, the analysis also demonstrated a significant dependence on the renormalization scale, with fluctuations of a few tens of MeV, which are of a similar magnitude to the expected precision of future collider experiments. \\ Several research perspectives are currently being considered for future investigations. First, it is planned to incorporate the newly derived momentum-dependent corrections into the complex mass pole of the Higgs propagator and examine their numerical impact on the theoretical prediction, as well as assess the perturbative stability of the $\overline{MS}$ renormalization scheme for electroweak precision observables. Furthermore, a comprehensive numerical analysis considering different renormalization prescriptions for the top quark mass will be explored. In addition, efforts will be made to extend the developed computational routines to incorporate quantum corrections to the masses of the SM gauge bosons, specifically $M_Z$ and $M_W$, at the same perturbative order as considered in this study. Moreover, there is ongoing consideration of including momentum-dependent Higgs self-energies at order $y_t^6$ originating from the stop sector of the MSSM within the Dimensional Reduction scheme \cite{DREG4}. \\ It is important to note that in MSSM scenarios, the theoretical uncertainties typically range from $1$ to $5$ GeV, which is an order of magnitude larger than the experimental error in $M_h$. Consequently, calculating the missing higher-order corrections becomes crucial. However, it should be acknowledged that incorporating at least one additional scale, namely the SUSY scale, is necessary for these calculations. Therefore, it is recognized that exploring alternative approaches to the integration-by-parts reductions will be essential in addressing the challenge posed by the large divergent coefficients of the master integrals. It is worth mentioning that this consideration holds true for higher-order perturbative calculations involving an arbitrary number of energy scales.                           
\appendix
\section{Sample of 3-loop Self-Energy Amplitudes} \label{App:A}
In this appendix, we present additional simplified analytical expressions for the different families of the amplitudes $\mathcal{A}_{2}$, $\mathcal{A}_{4}$, and $\mathcal{A}_{6}$. These expressions were obtained by carefully selecting a basis of master integrals and are utilized in calculating the three-loop corrections to the SM Higgs boson mass at order $y_t^6$ (see Table~\ref{tab:IntFam}). To access the complete list of amplitudes, please refer to the following link:  \\ \\ \url{https://github.com/fisicateoricaUDP/HiggsSM}. 
\begin{align}
\mathcal{A}_{2}^{^{\{12378\}}} & =y_{t}^{6}\left[t\,\left(-4G_{_{\{0,2,1,1,0,0,1,1,0\}}}+4G_{_{\{1,0,1,1,0,1,1,1,0\}}}\right.\right. \nonumber  \\
 & \qquad\quad-8G_{_{\{1,1,1,0,0,1,1,1,0\}}}+4G_{_{\{1,1,1,1,-1,1,1,1,0\}}} \nonumber \\
 & \qquad\quad-4G_{_{\{1,1,1,1,0,0,1,1,0\}}}+4G_{_{\{1,1,1,1,0,1,0,1,0\}}} \nonumber  \\
 & \qquad\quad+4G_{_{\{1,1,1,1,0,1,1,0,0\}}}-2G_{_{\{1,1,1,1,0,1,1,1,-1\}}} \nonumber \\
 & \qquad\quad+8G_{_{\{1,2,1,0,0,0,1,1,0\}}}-4G_{_{\{1,2,1,1,0,0,0,1,0\}}} \nonumber \\
 & \qquad\quad\left.-4G_{_{\{1,2,1,1,0,0,1,0,0\}}}\right) \nonumber \\
 & \qquad+t^{2}\,\left(8G_{_{\{0,2,1,1,0,1,1,1,0\}}}+18G_{_{\{1,1,1,1,0,1,1,1,0\}}}\right. \nonumber \\
 & \qquad\quad-16G_{_{\{1,2,1,0,0,1,1,1,0\}}}-16G_{_{\{1,2,1,1,0,0,1,1,0\}}} \nonumber \\
 & \qquad\quad\left.+8G_{_{\{1,2,1,1,0,1,0,1,0\}}}+8G_{_{\{1,2,1,1,0,1,1,0,0\}}}\right) \nonumber \\
 & \qquad\left.+t^{3}\,32G_{_{\{1,2,1,1,0,1,1,1,0\}}}\right] \nonumber \\
 & +sy_{t}^{6}\left[t\,\left(-2G_{_{\{1,1,1,1,0,1,1,1,0\}}}+4G_{_{\{1,2,1,1,0,0,1,1,0\}}}\right)\right. \nonumber \\
 & \qquad\left.-t^{2}\,8G_{_{\{1,2,1,1,0,1,1,1,0\}}}\right]
\end{align} 
\begin{align}
\mathcal{A}_{2}^{^{\{1278\}}} & =y_{t}^{6}\left[t\,\left(2G_{_{\{1,0,1,1,0,1,1,1,0\}}}-4G_{_{\{1,1,1,0,0,1,1,1,0\}}}\right.\right. \nonumber \\
 & \qquad\quad+2G_{_{\{1,1,1,1,-1,1,1,1,0\}}}+2G_{_{\{1,1,1,1,0,1,0,1,0\}}} \nonumber \\
 & \qquad\quad\left.+2G_{_{\{1,1,1,1,0,1,1,0,0\}}}-G_{_{\{1,1,1,1,0,1,1,1,-1\}}}\right) \nonumber \\
 & \qquad + t^{2}\,\left(-2G_{_{\{0,2,1,1,0,1,1,1,0\}}}-8G_{_{\{1,1,1,1,0,1,1,1,0\}}}\right. \nonumber \\
 & \qquad\quad+4G_{_{\{1,2,1,0,0,1,1,1,0\}}}-2G_{_{\{1,2,1,1,0,1,0,1,0\}}} \nonumber \\
 & \qquad\left.\left.-2G_{_{\{1,2,1,1,0,1,1,0,0\}}}\right)+t^{3}\,8G_{_{\{1,2,1,1,0,1,1,1,0\}}}\right] \nonumber \\
 & +sy_{t}^{6}\left[-t\, G_{_{\{1,1,1,1,0,1,1,1,0\}}}\right. \nonumber \\
 & \qquad \left.-t^{2}\, 2G_{_{\{1,2,1,1,0,1,1,1,0\}}}\right]
\end{align}
\vspace{-5mm}
\begin{align}
\mathcal{A}_{4}^{^{\{24589\}}} & =y_{t}^{6}\left[t\,\left(2G_{_{\{-1,1,0,1,1,1,1,1,1\}}}+2G_{_{\{0,0,0,0,1,1,1,2,1\}}}\right.\right. \nonumber \\
 & \qquad\quad+2G_{_{\{0,0,0,1,0,1,1,2,1\}}}-4G_{_{\{0,0,0,1,1,0,1,2,1\}}} \nonumber \\
 & \qquad\quad-4G_{_{\{0,0,0,1,1,1,0,2,1\}}}+2G_{_{\{0,0,0,1,1,1,1,1,1\}}} \nonumber \\
 & \qquad\quad+2G_{_{\{0,0,0,1,1,1,1,2,0\}}}-2G_{_{\{0,1,-1,1,1,1,1,1,1\}}} \nonumber \\
 & \qquad\quad+4G_{_{\{0,1,0,1,0,1,1,1,1\}}}-4G_{_{\{0,1,0,1,1,0,1,1,1\}}} \nonumber \\
 & \qquad\quad-6G_{_{\{0,1,0,1,1,1,0,1,1\}}}+4G_{_{\{0,1,0,1,1,1,1,0,1\}}} \nonumber \\
 & \qquad\quad\left.+2G_{_{\{0,1,0,1,1,1,1,1,0\}}}\right) \nonumber \\
 & \qquad+t^{2}\,\left(8G_{_{\{0,0,0,1,1,1,1,2,1\}}}+8G_{_{\{0,1,0,0,1,1,1,2,1\}}}\right. \nonumber \\
 & \qquad\quad+8G_{_{\{0,1,0,1,0,1,1,2,1\}}}-16G_{_{\{0,1,0,1,1,0,1,2,1\}}} \nonumber \\
 & \qquad\quad-16G_{_{\{0,1,0,1,1,1,0,2,1\}}}+18G_{_{\{0,1,0,1,1,1,1,1,1\}}} \nonumber \\
 & \qquad\quad\left.+8G_{_{\{0,1,0,1,1,1,1,2,0\}}}\right) \nonumber \\
 & \qquad\left.+t^{3}\,32G_{_{\{0,1,0,1,1,1,1,2,1\}}}\right] \nonumber \\
 & +sy_{t}^{6}\left[t\,\left(-2G_{_{\{0,1,0,0,1,1,1,2,1\}}}-2G_{_{\{0,1,0,1,0,1,1,2,1\}}}\right.\right. \nonumber \\
 & \qquad\quad+4G_{_{\{0,1,0,1,1,0,1,2,1\}}}+4G_{_{\{0,1,0,1,1,1,0,2,1\}}} \nonumber \\
 & \qquad\quad\left.-2G_{_{\{0,1,0,1,1,1,1,1,1\}}}-2G_{_{\{0,1,0,1,1,1,1,2,0\}}}\right) \nonumber \\
 & \qquad\left.+t^{2}\,8G_{_{\{0,1,0,1,1,1,1,2,1\}}}\right]
\end{align}
\begin{align}
\mathcal{A}_{6}^{^{\{125678\}}} & =y_{t}^{6}\left[t\,\left(2G_{_{\{0,1,0,1,1,1,1,1,0\}}}-2G_{_{\{0,1,0,2,0,1,1,1,0\}}}\right.\right. \nonumber \\
 & \qquad\quad-2G_{_{\{0,1,0,2,1,0,1,1,0\}}}-2G_{_{\{0,1,0,2,1,1,1,0,0\}}} \nonumber \\
 & \qquad\quad+2G_{_{\{1,0,0,1,1,1,1,1,0\}}}-2G_{_{\{1,0,0,2,0,1,1,1,0\}}} \nonumber \\
 & \qquad\quad-2G_{_{\{1,0,0,2,1,0,1,1,0\}}}-2G_{_{\{1,0,0,2,1,1,0,1,0\}}} \nonumber \\
 & \qquad\quad-4G_{_{\{1,1,0,0,1,1,1,1,0\}}}+4G_{_{\{1,1,0,1,0,1,1,1,0\}}} \nonumber \\
 & \qquad\quad+4G_{_{\{1,1,0,1,1,0,1,1,0\}}}+2G_{_{\{1,1,0,1,1,1,0,1,0\}}} \nonumber \\
 & \qquad\quad+2G_{_{\{1,1,0,1,1,1,1,0,0\}}}-2G_{_{\{1,1,0,2,0,1,0,1,0\}}} \nonumber \\
 & \qquad\quad-2G_{_{\{1,1,0,2,0,1,1,0,0\}}}-2G_{_{\{1,1,0,2,1,0,0,1,0\}}} \nonumber \\
 & \qquad\quad\left.-2G_{_{\{1,1,0,2,1,0,1,0,0\}}}\right) \nonumber \\
 & \qquad+t^{2}\,\left(-8G_{_{\{0,1,0,2,1,1,1,1,0\}}}-8G_{_{\{1,0,0,2,1,1,1,1,0\}}}\right. \nonumber \\
 & \qquad\quad+16G_{_{\{1,1,0,1,1,1,1,1,0\}}}-8G_{_{\{1,1,0,2,0,1,1,1,0\}}} \nonumber \\
 & \qquad\quad-8G_{_{\{1,1,0,2,1,0,1,1,0\}}}-8G_{_{\{1,1,0,2,1,1,0,1,0\}}} \nonumber \\
 & \qquad\quad\left.-8G_{_{\{1,1,0,2,1,1,1,0,0\}}}\right) \nonumber \\
 & \qquad\left.-t^{3}\,32G_{_{\{1,1,0,2,1,1,1,1,0\}}}\right] \nonumber \\
 & +sy_{t}^{6}\left[t\,\left(2G_{_{\{1,1,0,2,0,1,1,1,0\}}}+2G_{_{\{1,1,0,2,1,0,1,1,0\}}}\right)\right. \nonumber \\
 & \qquad\left.+t^{2}\,8G_{_{\{1,1,0,2,1,1,1,1,0\}}}\right]
\end{align}

\end{document}